\documentclass[twocolumn,showpacs]{revtex4}
\usepackage{epsfig}
\begin{document}

\title{
Mode structure and photon number correlations in squeezed quantum pulses
}
\author{T. Opatrn\'{y}} 
\affiliation{
Physikalisches Institut/Optik, Friedrich-Alexander University
Erlangen-N\"{u}rnberg, Germany
}
\affiliation{
Department of Physics, Texas A\&M University,
College Station, TX 77843-4242}
\affiliation{Department of Theoretical Physics, Palack\'{y} University,
Olomouc, Czech Republic}
\author{N. Korolkova}
\affiliation{
Physikalisches Institut/Optik, Friedrich-Alexander University
Erlangen-N\"{u}rnberg, Germany
}
\author{G. Leuchs} 
\affiliation{
Physikalisches Institut/Optik, Friedrich-Alexander University
Erlangen-N\"{u}rnberg, Germany
}

\date{\today}

\begin{abstract}
The question of efficient multimode description of optical pulses is studied.
We show that a relatively very small number of nonmonochromatic modes can be
sufficient for a complete quantum description of pulses with Gaussian
quadrature statistics.  For example, a three-mode description was enough to
reproduce the experimental data of photon number correlations in optical
solitons [S. Sp\"{a}lter et al., Phys. Rev. Lett. {\bf 81,} 786 (1998)]. This
approach is very useful for a detailed understanding of squeezing properties of
soliton pulses with the main potential for quantum communication with
continuous variables. We show how homodyne detection and/or measurements of
photon number correlations can be used to determine the quantum state of the
multi-mode field. We also discuss a possible way of physical separation of the
nonmonochromatic modes.
\end{abstract}


\maketitle

%%%%%%%%%%%%%%%%%%%%%%%%%%%%%%%%%%%%%%%%%%

%\begin{multicols}{2}

\section{Introduction}

For a complete quantum description of an optical pulse, one has to use a
multimode density matrix. The question is, how many modes are necessary for
such a description. If one  works with {\em monochromatic \/} modes, then an
infinite number of modes would be needed, which is impractical. On the other
hand, one can construct different sets of modes as linear combinations of the
monochromatic modes \cite{TG66}. These modes (which are however {\em
nonmonochromatic}) may be more suitable for the quantum description of pulses.
In particular, it would be useful to find such a modal structure  so
that the quantum state of the pulse can be described by the density operator
$\hat \varrho$ $=$ $\hat \varrho_{\rm exc}\otimes \hat \varrho_{\rm vac}$,
where $\hat \varrho_{\rm exc}$ is some nontrivial density operator of a few
modes and  $\hat \varrho_{\rm vac}$ is the vacuum state density operator of all
the remaining modes. Thus, we could work with a concise description  of the
pulse by means of  $\hat \varrho_{\rm exc}$.

Our work is motivated by the question how to completely describe the quantum
state of solitons in optical fibers.  Such soliton pulses are used in various
schemes of quantum information processing (for a review see, e.g.,
\cite{Natasa}). Can a single-mode description of a soliton pulse be sufficient?
When  explaining squeezing in the Kerr medium, one sometimes plots a picture of
phase space where a quantum uncertainty circle is deformed into an ellipse---a
description which clearly corresponds to a single mode situation. This approach
is useful to qualitatively understand the generation of continuous variables
entanglement as in \cite{Silberhorn}, but do we  use all the quantum features
of a given pulse if we treat it as a single mode object? Should one, instead,
work with many more modes of a pulse in the hope that  each of them can become
a useful resource for  quantum communication? Multi-mode correlations of photon
numbers in a soliton were observed  \cite{SKKSL98}.  To numerically calculate
such correlations, quantum variables of a soliton were treated on a position
grid of typically $\sim 10^2 - 10^3$ points  \cite{Spaelter-dis,SKW99}. Are
there hundreds of useful modes available in a pulse, or would a better choice
of the mode functions show that only a few modes  (perhaps just one?) are in
some nontrivial quantum state? Or are just four quantum operators introduced by
Haus and Lai \cite{HausLai} sufficient to describe all the relevant phenomena? 

Knowing the answer to the question of what is the complete quantum description
of a pulse would be very helpful in quantum information processing: one  could
fully utilize the squeezing and entanglement properties of our sources. After
forming a soliton pulse in a fiber, one can determine its  multimode quantum
state in the most comprehensive way. One can then optimize the medium
properties to achieve maximum squeezing, or maximum purity of an entangled
state. Working with pairs of correlated pulses, one could apply  a proper
measurement scheme and use entanglement criteria for multimode bipartite
Gaussian states \cite{EntCrit} to check wheteher the pulse pair is entangled or
separable. One can also better understand the influence of the medium on the
propagating pulse: provided that the output pulse is deformed, what happens
with the quantum information carried by the pulse? Is it washed out by
decoherence processes (or perhaps by an eavesdropper), or is it just unitarily
transformed into other modes of the same pulse?  A simple and correct
measurement and description of the multimode state is highly desirable.

This work is organized as follows. In Sec. \ref{S-modes} we introduce 
nonmonochromatic modes and deal with the transformations between different sets
of quadrature operators. Sec. \ref{S-photons} studies basic properties of the
photon statistics of multimode fields: mean photon numbers and covariances
between different modes. In Sec. \ref{S-variance} we discuss a homodyne scheme
for a complete determination of Gaussian states of nonmonochromatic multi-mode
fields. In Sec. \ref{S-comparison} we compare the  photon number squeezing
available  via applying a proper local oscillator modulation and via using
spectral filtering. Sec. \ref{S-modeselect} suggests a way for an optimal
selection of the nonmonochromatic mode functions. In Sec. \ref{Sec-Separ} we
discuss a possible way how to physically separate individual nonmonochromatic
modes of the pulse. Discussion and conclusion are presented in Sec.
\ref{S-conclusion}. Most of the mathematical details are discussed in
Appendixes \ref{App-simplify}--\ref{S-Elimin}, and in App. \ref{App-Haus} we
compare our approach to that of Haus and Lai \cite{HausLai}.

%%%%%%%%%%%%%%%%%%%%%%%%%%%%%%%%%%%%%%%%%%%%%%%%%%%%%%%%%%%%%%%%%%%%%%

\section{Nonmonochromatic modes}
\label{S-modes}

Under the term ``mode'' we understand a single degree of freedom of the
electromagnetic field; a mode can be described by a pair of bosonic operators.
It can be {\em monochromatic\/} (evolution described by a single frequency) or
{\em nonmonochromatic}. Any state of the field can be treated in different mode
decompositions. In a given decomposition the state is called {\em single
mode\/} if all modes except of one have vacuum statistics of their  operators,
in the opposite case the state is {\em multimode}. In this section we deal with
the transition between different mode decompositions. 

\subsection{Bosonic operators}

Let us assume polarized optical field propagating in a given direction. The
monochromatic modes of the field are denoted by the corresponding frequency
$\omega$, and the bosonic operators of these modes satisfy the commutation
relations
\begin{eqnarray}
[\hat a(\omega) , \hat a^{\dag}(\omega ')] = \delta (\omega-\omega'), \qquad
[\hat a (\omega) , \hat a (\omega' )] = 0.
\end{eqnarray}
Let us assume a complete orthonormal set of functions $f_k(\omega)$,
\begin{eqnarray}
 \label{fmfk}
 \int f_m^* (\omega) f_k (\omega) d\omega = \delta_{mk} ,\\
 \sum_{k} f_k^*(\omega ) f_k(\omega ) = \delta (\omega - \omega') .
 \label{fkfk}
\end{eqnarray}
We define a new set of operators $\hat b_k$ as
\begin{eqnarray}
 \hat b_k = \int f_k (\omega ) \hat a (\omega ) d\omega 
 \label{bUa}
\end{eqnarray} 
which satisfy the commutation relations
\begin{eqnarray}
[\hat b_k , \hat b_{k'}^{\dag}] = \delta_{kk'}, \qquad
[\hat b_k , \hat b_{k'}] = 0.
\end{eqnarray}
Thus, the functions $f_k(\omega)$
can be used to define a new set of {\em nonmonochromatic} modes.
The inverse transformation of the nonmonochromatic modes to the monochromatic
ones reads
\begin{eqnarray}
 \hat a(\omega ) = \sum_{k} f^*_k (\omega) \hat b_k .
 \label{aUb}
\end{eqnarray}

%%%%%%%

\subsection{Example}
\label{ssExample}

Let us consider a pulse with a normalized frequency envelope
$f(\omega)$; $\int |f(\omega)|^2 d\omega$ $=$ 1. Let us define 
an orthonormal system of mode
functions as in Eqs. (\ref{fmfk}) and
(\ref{fkfk}) with $f_1(\omega) \equiv f(\omega)$.
Let the quantum state of the first mode $\hat b_1$ be
a coherent state $|\beta \rangle_{1}$,  $\hat b_1 |\beta \rangle_{1}$ $=$
$\beta |\beta \rangle_{1}$,  and let all the other $b$-modes be in vacuum $|0
\rangle_{k}$,  $\hat b_k |0 \rangle_{k}$ $=$ 0, $k>1$. By means of the
transformation (\ref{aUb}) we find that using the monochromatic modes, the
quantum state is a multi-mode coherent state 
$\prod_{\omega}|f(\omega)\beta \rangle_{\omega}$.
Even though the single-mode and multimode
coherent states are related to the same object, i.e., a pulse in a coherent
state, the single-mode description is clearly more convenient for handling and
for understanding the field structure.
Of course, for other than coherent states the transformations of the state
in different mode-systems are not so straightforward, but still can
provide us with a very convenient way of the quantum state description.

%%%%%%

\subsection{Quadrature operators}

We define the hermitian 
quadrature operators $\hat  x (\omega)$, $\hat p (\omega)$ as
\begin{eqnarray}
 \hat  x (\omega) &=& \frac{1}{\sqrt{2}}\left( \hat a (\omega)
 + \hat a^{\dag} (\omega) \right), \\  
 \hat p(\omega) &=& \frac{1}{i\sqrt{2}}\left( \hat a (\omega) - \hat a^{\dag}
 (\omega )\right) ,
\end{eqnarray} 
and similarly the $b$-mode quadrature operators 
  $\hat  X_k$, $\hat  P_k$ as
\begin{eqnarray}
 \hat  X_k &=& \frac{1}{\sqrt{2}}
 \left( \hat  b_k + \hat  b_k^{\dag}\right) , \\  
 \hat  P_k &=& \frac{1}{i\sqrt{2}}\left(
 \hat  b_k - \hat b^{\dag}_k \right). 
\end{eqnarray}
These operators  obey the same commutation relations
as the quantum mechanical position and momentum operators with
$\hbar$ $=$ 1, i.e.,
\begin{eqnarray}
 \left[ \hat x(\omega),\hat p(\omega')\right] = i \delta (\omega - \omega'), \\
 \left[ \hat x (\omega),\hat x(\omega')\right]  =  
 \left[ \hat p (\omega),\hat p(\omega')\right] =0,
\end{eqnarray}
and
\begin{eqnarray}
 \left[\hat X_k,\hat P_{k'}\right] = i \delta_{kk'}, \\
 \left[ \hat X_k,\hat X_{k'}\right]  =  \left[ \hat P_k,\hat P_{k'}\right] =0.
\end{eqnarray}
The transformation between the two sets of quadrature operators can be written
in the form
\begin{eqnarray}
 \label{muZxi}
 \hat \mu_k &=& \sum_{\xi=x,p} \int Z^{\mu\xi}_k (\omega) \hat
 \xi (\omega) d\omega , \\
 \label{xiZmu}
 \hat \xi (\omega) &=& \sum_{\mu=X,P} \sum_{k} Z^{\mu\xi}_{k}
 (\omega) \hat \mu_k 
\end{eqnarray}
with $\xi = x,p$, and $\mu = X,P$, and the transformation matrix $Z$ is
\begin{eqnarray}
 Z^{Xx}_k(\omega) = Z^{Pp}_k (\omega) = \text{Re}\ f_k (\omega) 
 \label{ZXx}
 \\
 Z^{Xp}_k(\omega) = -Z^{Px}_k (\omega) = \text{Im}\ f_k (\omega) . 
\end{eqnarray}
As can be checked, matrix $Z$ satisfies the orthogonality relations
\begin{eqnarray}
 \label{normalZ1}
 \sum_{\xi} \int Z^{\mu\xi}_k (\omega ) Z^{\mu' \xi}_{k'}(\omega) d\omega
 = \delta_{\mu \mu'}\delta_{kk'} , \\
 \sum_{\mu} \sum_{k} Z^{\mu \xi}_k(\omega) Z^{\mu \xi'}_k (\omega')
 = \delta(\omega - \omega ') \delta_{\xi \xi'} .
 \label{completZ}
\end{eqnarray}
Sometimes it is useful to work with a discrete set of frequencies $\omega_j$
corresponding to frequency bins of width $\Delta \omega$. Quadrature 
$\hat \xi_j$ of the $j$th bin is obtained as
\begin{eqnarray}
 \label{xiZmu2}
 \hat \xi_j = \sum_{\mu=X,P} \sum_{k} Z^{\mu\xi}_{kj}
 \hat \mu_k , 
\end{eqnarray}
where the transformation matrix elements $Z^{\mu\xi}_{kj}$ are
\begin{eqnarray}
 Z^{\mu\xi}_{kj} \equiv Z^{\mu\xi}_k (\omega_j ) \sqrt{\Delta \omega} .
 \label{Zmuxikj}
\end{eqnarray}
For brevity, we can join the quadratures $\hat x(\omega)$ and $\hat p(\omega)$
into a single vector $\hat \xi$, and similarly the quadratures $\hat X_k$ and
$\hat P_k$ into a single vector $\hat \mu$, so that the transformations
(\ref{muZxi}) and (\ref{xiZmu}) [or (\ref{xiZmu2})]
are written in the matrix multiplication form
\begin{eqnarray}
 \hat \mu = Z \hat \xi, \qquad \hat \xi = Z^{T} \hat \mu .
 \label{matrixZ}
\end{eqnarray}
Here the superscript $T$ means matrix transposition, and the rows of
the matrix $Z$ are represented by indices $\mu$ and $k$, and the columns
by the indices $\xi$ and $\omega$ (or $j$).

%%%%%%%%%%%%%%%%%%%%

\subsection{Mean quadratures and variances}

Important properties of multimode quantum states are given by the vectors of
the quadrature mean values, $\overline{\xi}_j$ 
$\equiv$ Tr$[\hat \varrho \hat 
\xi_j]$ and $\overline{\mu}_k$ $\equiv$ Tr$[\hat \varrho \hat \mu_k]$,
and by the variance matrices $V$ and $V'$, given as
\begin{eqnarray}
  V_{\mu k, \mu' k'} &=& \frac{1}{2}\left\langle 
 \left\{ \left( \mu _k - \overline{\mu}_k\right) ,
 \left( \mu'_{k'} - \overline{\mu'}_{k'}\right)  \right\}
 \right\rangle  ,
 \\
 V'_{\xi j,\xi' j'} &=& 
  \frac{1}{2}\left\langle \left\{ \left(
 \xi_j - \overline{\xi} _j\right),
 \left( \xi' _{j'} - \overline{\xi '} _{j'}\right) \right\} 
 \right\rangle 
\end{eqnarray}
where $\{ \dots \}$ stands for the anticommutator, $\{ \hat A, \hat B\}$
$\equiv$ $\hat A \hat B$ $+$ $\hat B \hat A$, and $\langle \dots \rangle$
represents averaging $\langle \hat A \rangle$ $\equiv$ Tr$(\hat \varrho \hat A)$.
The relationship between
these quantities can be written in the matrix multiplication
form as 
\begin{eqnarray}
 \overline{\mu} &=& Z \overline{ \xi } , 
 \label{trans1} 
 \\
 \overline{ \xi } &=& Z^{T} \overline{\bf \mu } , 
 \label{trans2} 
 \\  
 V &=& Z V' Z^{T} , 
 \label{trans3} 
 \\
 V' &=& Z^{T} V Z .
 \label{trans4}  
\end{eqnarray}

In general, to transform between $\overline {\mu}$,  $V$, and   $\overline
{\xi}$, $V'$, one has to know the complete transformation matrix $Z$, i.e., the
whole set of mode functions $f_k(\omega)$.  However, if we assume that only a
few nonmonochromatic modes are excited while the rest is in vacuum state, then
the explicit form of the empty mode functions does not play any role. (It was
also illustrated in the Example \ref{ssExample} where only one mode function
$f(\omega)$ was enough to transform the state into the multi-mode case.) This
can be used for a simplified calculation of the transformed quantities, as
discussed in App. \ref{App-simplify}.

%%%%%%%%%%%%%%%%%%%%%%%%%%%%%%%%%%%%%%%%%%%%%%%%%%%%%%%%%%%%%%%%

\section{Photon statistics of multimode Gaussian states}
\label{S-photons}

Among all possible states, the class of Gaussian states is one of the most
important---both from the theoretical and experimental point of view. The basic
property of these states is that their  Wigner function (as well as the
$Q$-function or  the characteristic function) have Gaussian form (see, e.g.,
\cite{SMD94,EntCrit,Reconstr}).   Examples of Gaussian states are coherent
states, squeezed coherent states, thermal states, and squeezed thermal states. 
Gaussian states are typically observed in most experiments with optical pulses.
In this paper we confine ourselves to Gaussian states; the main advantage  used
here is that a Gaussian state remains Gaussian in any mode decomposition and a
relatively small number of parameters is necessary for its description. 

\subsection{Mean photon numbers and variances}

\begin{figure}[htb]
\noindent
\centerline{\epsfig{figure=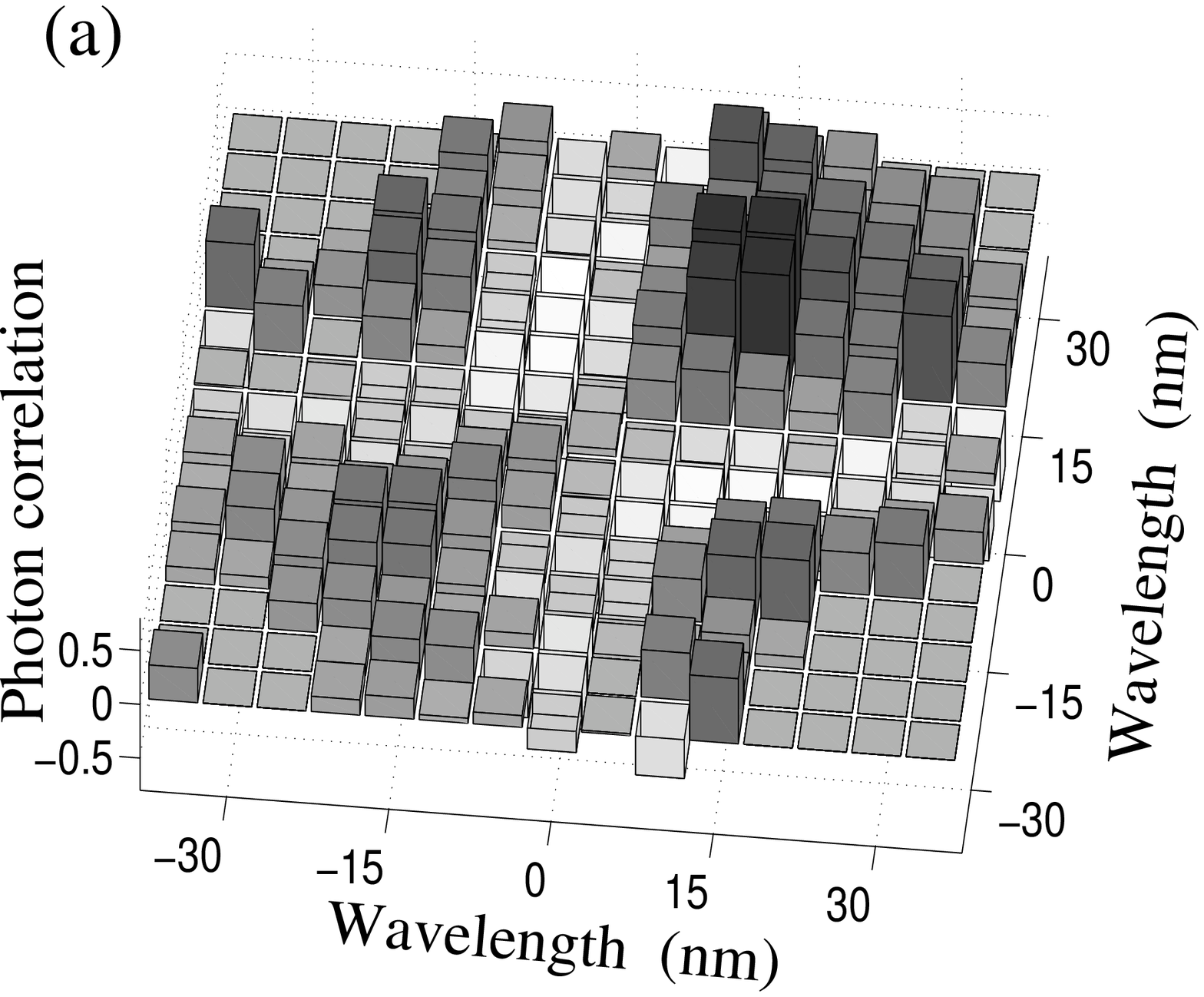,width=0.8\linewidth}}
\centerline{\epsfig{figure=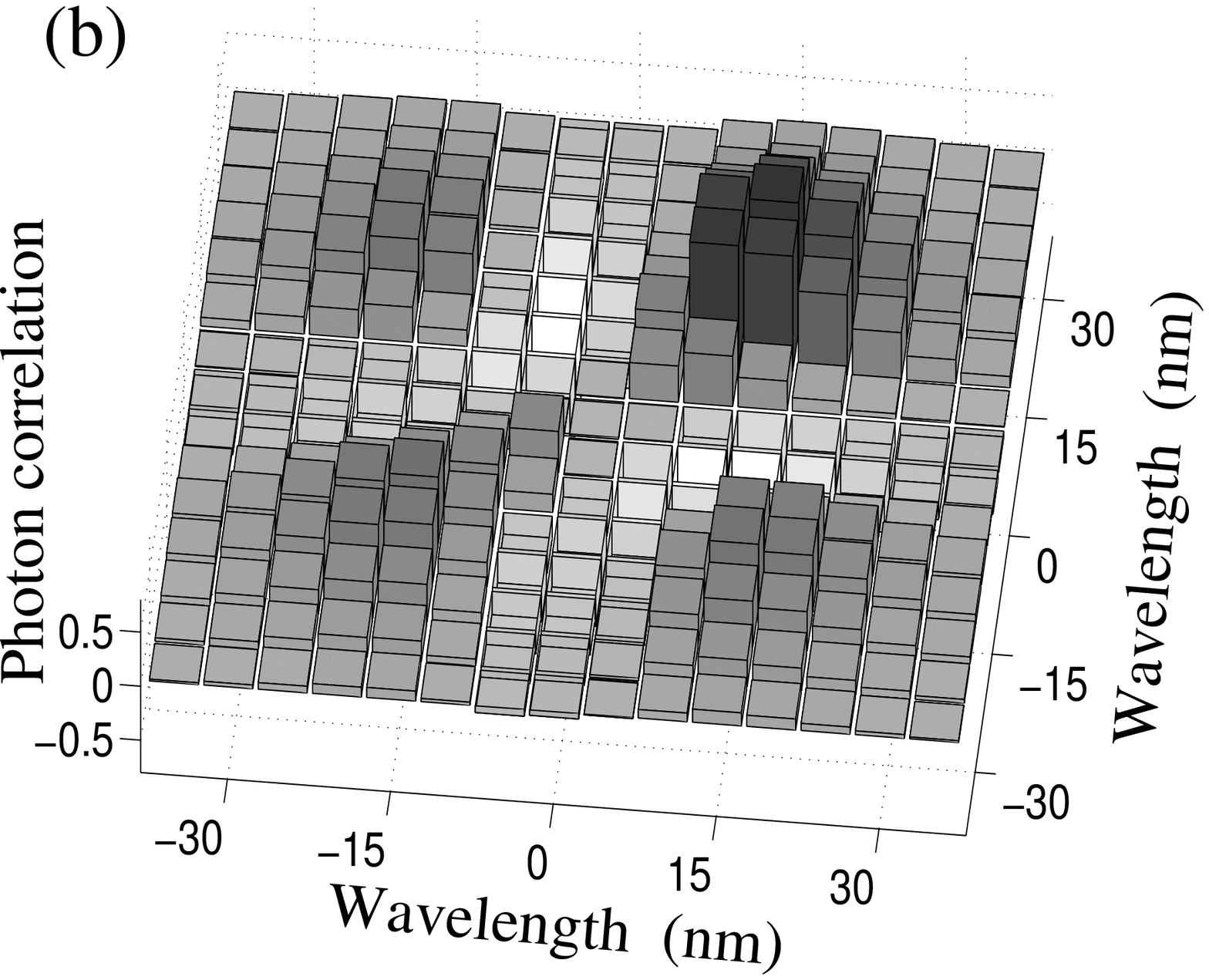,width=0.8\linewidth}}
\caption{ 
Photon correlation matrix $C^{\text{(n)}}_{kl}$ as a
function of the wavelengths of
the pulse spectrum. Results measured  in \cite{SKKSL98} 
with omitted elements with very large fluctuations ($> 0.8$) $(a)$, 
and the best fit using a reconstructed three-mode variance matrix $V_{Xk,Xk'}$ 
(see Sec. \ref{Subs-exper}) $(b)$.
}
\label{f-8}
\end{figure}

Let us first study the relation between quadrature moments and photon number
moments in discrete modes.
The photon number operator in $k$th mode $\hat n_k$ can be expressed by means of
the quadrature operators as
\begin{eqnarray}
 \hat n_k = \frac{1}{2} \left(\hat x_k^2 + \hat p_k^2 -1 \right) .
\end{eqnarray}
The mean photon number in $k$th mode is thus
\begin{eqnarray}
 \langle  n_k \rangle = \frac{1}{2} \left( \langle \hat x_k^2 \rangle
 + \langle\hat  p_k^2 \rangle - 1\right) ,
 \label{mfn1}
\end{eqnarray}
and the mean product of photon numbers  is
\begin{eqnarray}
 \langle  n_k  n_l \rangle = 
 \frac{1}{4} \left( \left\langle \hat x_k^2 \hat x_l^2
 \right\rangle + \left\langle \hat x_k^2 \hat p_l^2 \right\rangle +
 \left\langle \hat p_k^2 \hat p_l^2 \right\rangle 
 + \left\langle \hat  p_k^2 \hat x_l^2 \right\rangle  \right.
 \nonumber \\
 \left. - \left\langle \hat x_k^2  \right\rangle 
  - \left\langle \hat p_k^2  \right\rangle 
  - \left\langle \hat x_l^2  \right\rangle 
  - \left\langle \hat p_l^2  \right\rangle   
  + 1
 \right) .
 \label{mfn3}
\end{eqnarray}
As discussed in App. \ref{App-Gauss},
for Gaussian states all the quadrature moments on right hand sides of
Eqs. (\ref{mfn1}) -- (\ref{mfn3}) can be expressed 
using the quadrature means and
variances as in Eqs. (\ref{ximom2}) -- (\ref{ximom6}). 
The resulting photon number moments can be used to
calculate the photon number covariances
\begin{eqnarray}
 {\rm cov}\ (n_k, n_l) \equiv  \langle  n_k  n_l \rangle 
 -  \langle  n_k \rangle  \langle  n_l \rangle ,
 \label{covar}
\end{eqnarray}
which can be written in terms of the quadrature means and variances as
\begin{eqnarray}
{\rm cov}\ (n_k, n_l) = \bar x_k \bar x_l V'_{xk,xl} +
 \bar x_k \bar p_{l}V'_{xk,pl} 
 \nonumber \\ 
 + \bar p_{k} \bar x_{l}V'_{pk,xl} 
 + \bar p_{k} \bar p_{l}V'_{pk,pl} 
 \nonumber \\  
 +\frac{1}{2} \left(  V^{\prime 2}_{xk,xl} +   V^{\prime 2}_{pk,xl} +
 V^{\prime 2}_{xk,pl} +   V^{\prime 2}_{pk,pl} 
 \right)  - \frac{\delta_{kl}}{4},
 \label{covarnxp}
\end{eqnarray}
with the special case $\Delta n_{k}^2$ $\equiv$ cov$(n_k,n_k)$.
A very important role is played by the so called normally ordered covariance
\begin{eqnarray}
 C_{kl} \equiv 
 \langle : \Delta \hat n_k \Delta \hat n_l : \rangle   =  \text{cov} (n_k,n_l)   -  
 \delta _{kl}\langle n_k \rangle . 
 \label{normcov}
\end{eqnarray}
The normally ordered
covariance is used to define
the normalized
correlation matrix as in \cite{SKKSL98},
\begin{eqnarray}
  C^{\text{(n)}}_{kl} & \equiv & 
 \frac{\langle : \Delta \hat n_k \Delta \hat n_l : \rangle}
 {\sqrt{\Delta  n_k^2 \Delta  n_l^2 }} 
 = \frac{C_{kl}}{\sqrt{\Delta n_k^2 \Delta n_l^2}}.
 \label{correl-off} 
\end{eqnarray}
Values of the correlation matrix measured in \cite{SKKSL98} are
shown in Fig. \ref{f-8}a.
In the limit of continuous frequency modes, one can 
define photon number density $\langle n(\omega) \rangle$, variance density
$\langle \Delta n^2(\omega) \rangle$  $= \langle n(\omega) \rangle$,
and the normally ordered covariance
$C(\omega,\omega')$ with its normalized version 
$C^{\text{(n)}}(\omega,\omega')$.
Since a Gaussian state remains Gaussian in any mode decomposition, and since
the parameters of a Gaussian state $\bar \xi$ and $V$ can easily be transformed
from one mode decomposition into another, 
one can also calculate the photon number
mean values and correlations in arbitrary mode decomposition.

%%%%%%%%%%%%%%%%%%  F I G U R E %%%%%%%%%%%%%%%%%%%%%%
%%%%%%%%%%%%%%%%%%%%%%%%%%%%%%%%%%%%%%%%%%%%%%%%%%%%%%%

%%%%%%%%%%%%%%%%%%%

\subsection{Photon correlations in multimode pulses}

\label{Subsec-phot-stat}

If we try to reproduce the experimental results of
\cite{SKKSL98} as truly as possible  with as few modes as possible,
the first attempt would be to start with just a single nonmonochromatic mode.
However, as shown in Appendix \ref{App-singlemode}, the covariance matrix
cov$(n_k,n_l)$ of such a field would have the same sign for all values $k$ and
$l$. The sign would be positive if the single-mode state is super-Poissonian, or
negative if the single-mode state is sub-Poissonian. On the other hand, the
covariance matrix measured in \cite{SKKSL98} contains 
both positive and negative elements. Therefore, the quantum state of the soliton
pulses measured in  \cite{SKKSL98} {\em cannot be described as a single
nonmonochromatic mode.\/}

Exact analytical expressions are rather lengthy if more than one mode are
excited. However, relatively simple relations can be obtained if the following 
requirements are met: $(i)$  The coherent amplitude of at least one of the
modes is much greater than any of the variance matrix elements,  $(ii)$ only
the $X$-quadratures have non-zero mean values, i.e., $\bar P_k = 0$ for all
$k$,  and $(iii)$ the mode functions are real. While the condition $(i)$ is
generally valid for all strong pulses, the conditions $(ii)$ and $(iii)$ are
related to our choice of mode functions
and their generalization is straightforward.  
In Eq. (\ref{covarnxp}) the terms
containing $\bar p_{k,l}$ vanish 
and the first term on the right is dominant;
thus Eq. (\ref{covarnxp}) becomes
\begin{eqnarray}
 \text{cov}\ (n_k,n_l) \approx \bar x_k \bar x_l V'_{xk,xl}.
\end{eqnarray}
To calculate this quantity using the nonmonochromatic modes, we write
\begin{eqnarray}
 \bar x_{k} = \sum_{n\leq N} f_{n}(\omega_k) \bar X_{n} \sqrt{\Delta \omega}
\end{eqnarray}
(see Eqs. (\ref{trans2}), (\ref{ZXx}), and  (\ref{Zmuxikj})),
and for $V'_{xk,xl}$ we use Eq. (\ref{simplVI}) from Appendix
\ref{App-simplify}.
The mean photon number is
\begin{eqnarray}
 \langle n_k \rangle \approx \frac{1}{2} \sum_{m,n\leq N} f_{m}(\omega_k)
 f_{n}(\omega_k) \bar X_m \bar X_n \Delta \omega ,
 \label{meannk}
\end{eqnarray}
using Eqs. (\ref{mfn1}), (\ref{ximom2}), and the conditions $(i)$--$(iii)$.
The normally ordered covariance (\ref{normcov}) thus becomes
\begin{eqnarray}
 C_{kl} = \sum_{m,n \leq N} f_{m}(\omega_k) f_{n}(\omega_l)
 \left( V_{Xm,Xn} - \frac{\delta_{mn}}{2} \right) 
 \nonumber \\
 \times \sum_{m',n'\leq N} f_{m'}(\omega_k) f_{n'}(\omega_l) \bar X_{m'} \bar
 X_{n'} \Delta \omega ^2 .
\end{eqnarray}
Since in very narrow frequency bins $\Delta \omega$, the mean photon number and 
photon number variance are approximately equal, $\langle \Delta n^2_k \rangle$
$\approx$ $\langle n_k \rangle$ $\ll 1$, Eq. (\ref{meannk}) can be used also to
obtain $\langle \Delta n^2_k \rangle$, and the normalized correlation matrix
(\ref{correl-off}) becomes
\begin{eqnarray}
 C^{\text{(n)}}_{kl} \!=\! 2\! \sum_{m,n\leq N} f_m(\omega_k) f_n(\omega_l)
 \left( V_{Xm,Xn} -\frac{\delta_{mn}}{2} \right) \Delta \omega.
 \label{corm-disc}
\end{eqnarray}
Taking $\Delta \omega \to 0$, we can work with the continuous quantity
\begin{eqnarray}
 \label{cnorm-cont}
 C^{\text{(n)}}(\omega,\omega') \! = \! 2 \! \sum_{m,n\leq N} f_{m}(\omega)
 f_{n}(\omega') \left( V_{Xm,Xn} - \! \frac{\delta_{mn}}{2} 
 \right)  . \nonumber \\
\end{eqnarray}
In Eq. (\ref{cnorm-cont}) only the 
$X$-elements of the quadrature variances $V$ occur.
In other words, 
the measured photon number covariances are only influenced by the
covariances of quadratures which are in phase with the quadrature of the
strongly excited modes.

%%%%%%%%%%%%%%%%%%

\subsection{Reconstruction of the quadrature variances from
the photon number covariances}

\label{Subs-reconstr}

In the preceding subsection we have seen how to calculate the 
normalized photon covariance from the quadrature variance matrix of the
nonmonochromatic modes. We can also consider an inverse problem.
Let us assume that the mode functions $f_k(\omega)$ are known.
Let us also assume that the normally ordered covariances 
$C(\omega,\omega')$ are measured as in
\cite{SKKSL98} so that the normalized covariance 
$C^{\text{(n)}}(\omega,\omega')$ can be determined.
As can be seen from Eq. (\ref{cnorm-cont}), this quantity 
 is a linear combination
of the products of the mode functions $f_k(\omega)$. The coefficients 
in the linear combination are elements of the
quadrature variance $V$.
Thus, to obtain the elements of $V$, one can use the orthonormality of the mode
functions $f_k(\omega)$ to invert the linear dependence. We find
\begin{eqnarray}
 V_{Xk,Xk'} = \frac{1}{2} \delta_{kk'} \nonumber \\
 + \frac{1}{2} \int \int C^{\text{(n)}}(\omega,\omega')
 f_{k}(\omega) f_{k'}(\omega') d\omega d\omega' .
 \label{VXXreconstr}
\end{eqnarray}
Thus, from the spectral covariances of the photon numbers one can reconstruct
the $N(N+1)/2$ elements of the quadrature variance matrix, out of the total
number of $N(2N+1)$ independent elements.

Since Eq. (\ref{VXXreconstr}) is linear, it allows for a direct estimation of
the reconstruction error (see, e.g., \cite{Reconstr}). 
If the normalized covariances were measured with 
precision $\Delta C^{\text{(n)}}(\omega,\omega')$, then the error in
reconstructing the elements of $V$ can be estimated as
\begin{eqnarray}
 \langle \Delta V^2_{Xk,Xk'} \rangle \approx {1\over 4} \int \int 
 \Delta C^{\text{(n)}2}(\omega,\omega') \nonumber \\
 \times
 f^2_k(\omega) f^2_{k'}(\omega') 
  d\omega d\omega' .
\end{eqnarray}
Note that here, for the sake of brevity,
we have assumed  uncorrelated errors in the elements of 
$C^{\text{(n)}}(\omega,\omega')$; derivation of a more general formula
taking into account correlated errors is straightforward.

Let us stress that the formulas derived in  here and in Sec.
\ref{Subsec-phot-stat}  are valid for very narrow frequency bins where $\langle
\Delta n^2_k \rangle$ $\approx$ $\langle n_k \rangle$ $\ll 1$.   However, in
real experiments, it is often necessary to work with wider bins in which the
photon statistics can differ from Poissonian so that using Eqs.
(\ref{cnorm-cont}) and (\ref{VXXreconstr})  could cause significant errors.
This situation can be easily taken into account by substituting the
proper expression for  $\langle \Delta n^2_k \rangle$ in the formulas
connecting the photon correlation matrix with the quadrature variance matrix.
The equations (\ref{cnorm-cont}) and (\ref{VXXreconstr}) then become slightly
more involved; since this generalization is straightforward, we do not include
the formulas in this text.

%%%%%%%%%%%%%%%%%%%

\subsection{Application to experimental data}

\label{Subs-exper}

\begin{figure}[tb]
\noindent
\centerline{\epsfig{figure=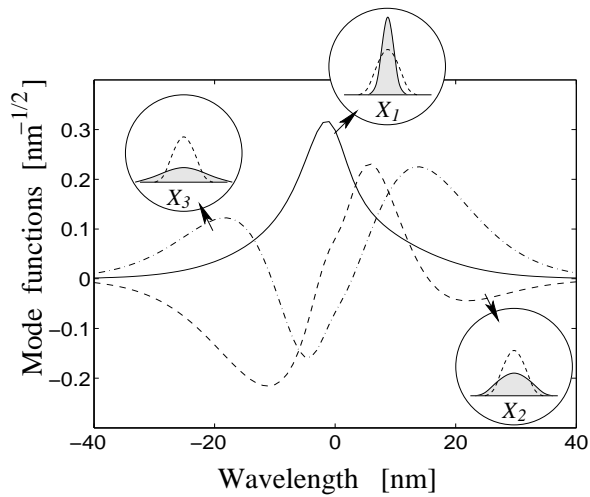,width=0.9\linewidth}}
\caption{ Mode functions for which the $V_{Xk,Xk'}$ matrix 
of the pulse measured in \cite{SKKSL98} is diagonal. The insets show the
fluctuations of the $X$-quadratures in comparison with the vacuum fluctuation
(dashed line). In the main figure, the full line corresponds to the mode
function $f_1$ with the squeezed quadrature, $V_{X1,X1} \approx 0.29$,
dashed line to $f_2$ with $V_{X2,X2}
\approx 1.39$, and the dash-dotted line to $f_3$
with $V_{X3,X3} \approx 2.69$.
}
\label{f-modes}
\end{figure}

To illustrate our method, we have used the experimental data obtained in
\cite{SKKSL98} (reproduced in Fig. \ref{f-8}a).  They were measured for
soliton pulses propagating in a 2.7~m optical fiber. We have used a preliminary
set of mode functions (starting with a sech function as the basic shape of the
soliton) and applied the reconstruction formula (\ref{VXXreconstr}). We have
diagonalized the resulting variance matrix $V_{Xk,Xk'}$ and found that only
three eigenvalues are substantially different from the vacuum value $1/2$. The
corresponding eigenvectors can be used to construct a new set of mode functions
(see Fig. \ref{f-modes}). In this set, the $X$-quadratures are independent of
each other. Thus, we have found that with respect to the quadratures which are 
in phase with the coherent amplitude ($X$-quadratures in our convention), the
pulse measured in \cite{SKKSL98} can effectively be described as a three mode
field, the quadrature variances being $V_{X1,X1} \approx 0.29$, $V_{X2,X2}
\approx 1.39$, and $V_{X3,X3} \approx 2.69$ with the off-diagonal elements
being zero. The smallest eigenvalue corresponds to squeezing $-2.35\ $dB.  The
reconstructed variance matrix $V_{Xk,Xk'}$ can be used to calculate back the
photon number correlation matrix $C^{\text{(n)}}_{kl}$; see Fig. \ref{f-8}b.

Since the photon number correlations were measured with limited precision, 
these results suffer from errors. To estimate the precision of our results, we
have Monte-Carlo generated 1000 ``experimental'' matrices $C^{\text{(n)}}_{kl}$
with elements fluctuating with errors given by the error estimates of the
original experiment.  The squeezing value then fluctuated between $-2.2\ $dB
and $-4.1\ $dB, with most results centered around $-3.1\ $dB. Even though the
error is too large to make a definite statement, these result suggest that
larger squeezing is available than the measured $\approx -2.5\ $dB of this
setup with spectral filtering. Let us note that the shape of the mode function
corresponding to the squeezed quadrature (full line in Fig. \ref{f-modes})
resembles the spectral filtering approach: the contribution from the middle of
the spectrum is enhanced while the outer parts are suppressed. The question of
photon number squeezing availability via spectral filtering and via local
oscillator functions is studied in more detail in Sec. \ref{S-comparison}.

%%%%%%%%%%%%%%%%%%  F I G U R E %%%%%%%%%%%%%%%%%%%%%%
%%%%%%%%%%%%%%%%%%%%%%%%%%%%%%%%%%%%%%%%%%%%%%%%%%%%%%%

\subsection{Conclusion}

We have seen that one mode  description of a pulse is not sufficient to explain
the experimental data of \cite{SKKSL98}, whereas a three mode description gives
a good agreement with the measurements.  However, we cannot conclude from this
that the pulse does not contain more than three modes. In other modes the 
$P$-quadratures  can be excited which do not influence the observed photon
statistics. Also, the measurement noise was too high so that weak excitation of
some additional modes might be undistinguished from the data noise backgraund.

It is interesting to compare our approach with that of Haus and Lai
\cite{HausLai} (see also \cite{Kaup90,Haus00,MK00}). In their case the quantum
field is decomposed into a ``soliton'' part and a ``continuum'' part. The
soliton field is described by four operators  $\Delta \hat n$, $\Delta \hat
\theta$, $\Delta \hat x$, and $\Delta \hat p$, related to the soliton energy,
phase, position, and velocity, respectively. As shown in App. \ref{App-Haus},
these operators can be expressed by our quadrature operators of four modes,
provided that the mode functions are properly selected. However, only two of
these operators ($\Delta \hat n$ and $\Delta \hat x$) related to two our modes
($f_1$ and $f_2$) influence the photon statistics. Thus, confining ourselves to
the four soliton operators of \cite{HausLai} would not be sufficient to
describe the observed phenomena. To apply the formalism of \cite{HausLai}, one
would have to work with the full set of the soliton and continuum operators.

%%%%%%%%%%%%%%%%%%%%%%%%%%%%%%%%%%%%%%%%%%%%%%%%%%%%%%%%%%%%%%%%%%%

\section{Complete determination of the variance matrix}
\label{S-variance}

Even though one can obtain full information about the $X$-quadratures from the
spectral correlations of photon numbers, one has no access to the variances
$V_{Pk,Pk'}$ and $V_{Xk,Pk'}$. To get full information on the multimode
Gaussian quantum state, one has to perform phase dependent measurements. 
Homodyne detection is one example of such a measurement; recently it was
studied how to apply homodyne detection scheme to the quantum state
reconstruction of a multimode optical field \cite{Reconstruc,OWV97}. Because we
confine ourselves to the Gaussian states, the task is easier than
reconstruction of a general quantum state.

\begin{figure}[!t]
\noindent
\centerline{\epsfig{figure=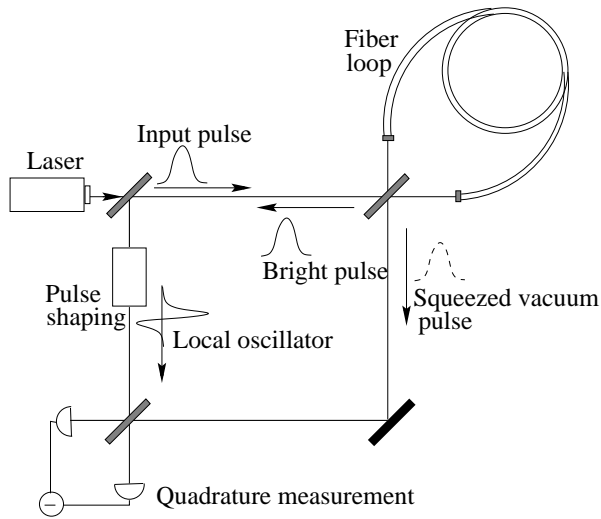,width=0.9\linewidth}}
\\[1ex]
\caption{ 
Scheme to measure all the elements of the quadrature variance matrix. The Sagnac
interferometer produces in one output squeezed vacuum and in the other output a
bright pulse corresponding to one of the mode functions (say $f_1(\omega)$). 
The 
pulse shaping device transforms the bright pulse envelope into different
combinations of the mode functions. Using the resulting bright pulse as a local
oscillator in a balanced homodyne detector, one can reconstruct the quadrature
variance matrix $V$.}
\label{f-loop}
\end{figure}

To find the variance matrix $V$, one can use the scheme of \cite{OWV97} with
the local oscillator pulses shaped to the form of weighted combinations of the
mode functions. It is very useful to subtract the coherent amplitude of the
pulse  by using a balanced Sagnac interferometer \cite{SH90} as in Fig.
\ref{f-loop}. Here, two counterpropagating identical squeezed pulses are
interfering at a 50\%/50\% beam splitter. In one of the outputs of the beam
splitter the pulses interfere constructively and form a bright pulse, which is
then used to form the local oscillator. In the other output where the pulses
interfere destructively,  squeezed vacuum (or a more general field with zero
mean amplitude) is formed. The squeezed vacuum pulse has the same quadrature
variance matrix $V$ as each of the two counterpropagating pulses.

The bright pulse is shaped interferometrically so that its envelope has the
form of different combinations of the mode functions $f_k(\omega)$. Such a
pulse is then used as a local oscillator in a balanced homodyne detector. Thus,
if the local oscillator is  $f_k(\omega)$, one can obtain the value of
$V_{Xk,Xk}$, whereas with the local oscillator  $if_k(\omega)$ one can obtain
$V_{Pk,Pk}$. Knowing these values and using the local oscillator pulse of the
form $f_k(\omega)+f_{k'}(\omega)$  one can obtain the value of $V_{Xk,Xk'}$, by
using the local oscillator form $i(f_k(\omega)+f_{k'}(\omega))$   one can
obtain the value of $V_{Pk,Pk'}$, and by using the form
$f_k(\omega)+if_{k'}(\omega)$ one can obtain the value of $V_{Xm,Pm'}$. 
Altogether, $N(2N+1)$ different forms of the local oscillator are sufficient to
obtain all the $N(2N+1)$ independent elements of the variance matrix $V$. Let
us note that to increase the precision of the measurements, one can increase
the number of different phases of the local oscillator; a Maximum-Likelihood
method for parameter estimations of a single mode Gaussian state using many
phases has been discussed in \cite{DAriano}.

%%%%%%%%%%%%%%%%%%  F I G U R E %%%%%%%%%%%%%%%%%%%%%%
%%%%%%%%%%%%%%%%%%%%%%%%%%%%%%%%%%%%%%%%%%%%%%%%%%%%%%%

Having found the variance matrix, one has all the information about the Gaussian
state. Then we can immediately see, e.g., what is the maximum available
squeezing of the state: it is the value corresponding to the minimum eigenvalue
of $V$. Typically, this value will be smaller than the minimum diagonal element
of $V$, which means that the optimum squeezing is shared among different modes.
For example, there is a quadrature squeezing in the fundamental mode $f_1$ due
to the Kerr effect (assuming mode functions as in App. \ref{App-Haus}), but the
$X_1$ quadrature is also correlated to the $X_3$ quadrature, because of the
correlation between the pulse width and energy. Therefore, better squeezing can
be expected  to occur in a combination of the modes 1 and 3 than in the
isolated mode 1. Thus, if the scheme is used to produce squeezing, the
measurement scheme can serve as a tool for a selection of the optimum local
oscillator.

One can also see how much are individual modes entangled with each other. To
take full advantage of this knowledge one has to be able to separate individual
modes from each other and distribute them among Alice, Bob, and other
entanglement consumers. This is, however, a rather nontrivial task; in Sec.
\ref{Sec-Separ} we will briefly mention a possible approach to its solution.

%%%%%%%%%%%%%%%%%%%%%%%%%%%%%%%%%%%%%%%%%%%%%%%%%%%%%%%%%

\section{Photon number squeezing via local oscillator modulation vs. spectral
filtering}

\label{S-comparison}

Let us assume that we want to prepare a pulse with the maximum photon number
squeezing, i.e., the photon number fluctuates as little as possible.
To quantify the photon number squeezing, one uses the Mandel $Q$-parameter
\cite{Mandel}
defined as
\begin{eqnarray}
 Q = \frac{\langle \Delta n^2 \rangle - \langle n \rangle}{\langle n \rangle},
 \label{MandelQ}
\end{eqnarray}
where $n$ refers to the photon number of the entire pulse. 
This quantity is negative for sub-Poissonian field and positive for
super-Poissonian fields.
It was suggested in
several works \cite{Friberg}
that spectral filtering of the pulse can lead to improvement of
the photon number squeezing by blocking frequency bands with 
correlated photon fluctuations and letting through frequency bands with
anticorrelated photon numbers. Here we show that a proper modulation of the
coherent amplitude (playing the role of the local oscillator) leads to the
optimum squeezing which cannot be overcome by the spectral filtering method.

%%%%%%%%%%%%%%%%%%%

\subsection{Local oscillator modulation}

Let us first assume that the quadrature variance matrix $V$ is fixed while we
can modulate the mean values of the quadratures $\bar X_n$ and $\bar P_n$; this
corresponds to the experimental situation as in Sec. \ref{S-variance} and Fig.
\ref{f-loop}. 
The photon number variance is
\begin{eqnarray}
 \langle \Delta n^2 \rangle = \sum_{k,l} \text{cov}\ (n_k,n_l).
\end{eqnarray}
Assuming as in Sec. \ref{Subsec-phot-stat} that the coherent amplitudes are much
greater than the variance matrix elements, and that the mode functions are real
(generalization to complex functions being straightforward),
we can express the photon number variance as
\begin{eqnarray}
  \langle \Delta n^2 \rangle = \sum_{m,n\leq N}
  \left( \bar X_m \bar X_n V_{Xm,Xn} \right.
  \nonumber \\
  \left. + 2 \bar X_m \bar P_n V_{Xm,Pn}
  + \bar P_m \bar P_n V_{Pm,Pn} \right).
\end{eqnarray}
The mean photon number is
\begin{eqnarray}
 \langle n \rangle = \sum_{k}\langle n_k \rangle =
 \frac{1}{2}\sum_{n\leq N}  \left( \bar X_n ^2 + \bar P_n^2 \right),
\end{eqnarray}
so that the $Q$-parameter of Eq. (\ref{MandelQ}) can be written,
after some algebra, as
\begin{eqnarray}
 Q = 2\sum_{m,n\leq N} \left[ 
 \tilde X_m \left( V_{Xm,Xn} -\frac{\delta_{mn}}{2}\right)
 \tilde X_n \right. \nonumber \\
 \left. + 2 \tilde X_m V_{Xm,Pn}\tilde P_n
 + \tilde P_m \left( V_{Pm,Pn} - \frac{\delta_{mn}}{2}\right) \tilde P_n 
 \right],
\end{eqnarray}
where
\begin{eqnarray}
 \tilde X_m \equiv \frac{\bar X_m}{\sqrt{\sum_{n\leq N} \left( 
 \bar X_n^2 + \bar P_n^2\right)}}, \\
 \tilde P_m \equiv \frac{\bar P_m}{\sqrt{\sum_{n\leq N} \left( 
 \bar X_n^2 + \bar P_n^2\right)}}.
\end{eqnarray}
Thus the $Q$-parameter can be expressed in the matrix multiplication form as 
\begin{eqnarray}
 Q = 2 \left( \tilde X^T, \tilde P^T \right) 
 \left(
 \begin{array}{cc}
  V_{XX} & V_{XP} \\
  V_{PX} & V_{PP} 
 \end{array}
 \right)
 \left(
 \begin{array}{c}
 \tilde X \\
 \tilde P
 \end{array}
 \right)
 -1,
\end{eqnarray}
where $\tilde X$ and $\tilde P$ are column vectors with the $\tilde X_m$ and
$\tilde P_m$ elements, and $V_{XX}$, $V_{XP}$, and $V_{PP}$ are the corresponding
submatrices of the variance matrix $V$. The variance matrix is multiplied from
the left and from the right with a unit vector, so that the $Q$-parameter is
limited by
\begin{eqnarray}
 2 \left( V^{\text{(min)}}-\frac{1}{2} \right) 
 \leq Q \leq 2 \left( V^{\text{(max)}}-\frac{1}{2}\right) ,
\end{eqnarray} 
where $V^{\text{(min)}}$ is the minimum eigenvalue of the variance matrix $V$
and $V^{\text{(max)}}$ is its maximum eigenvalue. 
The minimum value of $Q$ is reached if the vector of mean values of $X$ and
$P$ is the eigenvector of $V$ corresponding to its minimum
eigenvalue. Setting the components of this vector is possible using the scheme
of Fig. \ref{f-loop}.

%%%%%%%%%%%%%%%%%%%%

\subsection{Spectral filtering}

Let us assume a spectral filtering function $0\leq c(\omega)\leq 1$:
the frequency component is completely transmitted if $c(\omega)=1$ 
and it is completely blocked if $c(\omega) = 0$. This function
transforms the mean quadratures and variances as
\begin{eqnarray}
 \label{xfilt}
 \bar x_k^{f} &=& c(\omega_k) \bar x_k, \\
 \bar p_k^{f} &=& c(\omega_k) \bar p_k, \\
 \label{Vfxx}
 V^{\prime f}_{xk,xl} &=& c(\omega_k)V'_{xk,xl}c(\omega_l) + \frac{\delta_{kl}}
 {2} \left[ 1- c^2(\omega_k) \right] , \\
 \label{Vfpp}
 V^{\prime f}_{pk,pl} &=& c(\omega_k)V'_{pk,pl}c(\omega_l) + \frac{\delta_{kl}}
 {2} \left[ 1- c^2(\omega_k) \right] , \\
 \label{Vfxp}
 V^{\prime f}_{xk,pl} &=& c(\omega_k)V'_{xk,pl}c(\omega_l) . 
\end{eqnarray}
The $\delta_{kl}$ terms in Eqs. (\ref{Vfxx}) and (\ref{Vfpp}) follow from the
quantum mechanical nature of the quadratures:
partially blocking a frequency component means that the corresponding field is
mixed with vacuum. 

Eqs. (\ref{xfilt})--(\ref{Vfxp}) can be used to determine $\langle \Delta n^2
\rangle$ and $\langle n \rangle$ as in the preceding subsection, and thus 
to find
the $Q$-parameter of the filtered field. After a straightforward algebra, one
can express the new $Q$-parameter as
\begin{eqnarray}
  Q^{f} = 2\sum_{m,n\leq N} \left[ 
  \tilde X^{f}_m \left( V_{Xm,Xn} -\frac{\delta_{mn}}{2}\right)
  \tilde X^{f}_n \right. \nonumber \\
 \left.  + 2 \tilde X^{f}_m V_{Xm,Pn}\tilde P^{f}_n 
  + \tilde P^{f}_m \left( V_{Pm,Pn} - \frac{\delta_{mn}}{2}\right) 
  \tilde P^{f}_n 
  \right],
\end{eqnarray}
where
\begin{eqnarray}
 \label{tildeXf}
 \tilde X^{f}_m = 
 \frac{\sum_{n\leq N}c_{mn}\bar X_n}{\sqrt{\sum_{n,n'\leq N}
 c_{nn'}\left( \bar X_n \bar X_{n'} + \bar P_{n} \bar P_{n'}\right)}}, \\
 \label{tildePf}
 \tilde P^{f}_m = 
 \frac{\sum_{n\leq N}c_{mn}\bar P_n}{\sqrt{\sum_{n,n'\leq N}
 c_{nn'}\left( \bar X_n \bar X_{n'} + \bar P_{n} \bar P_{n'}\right)}} ,
\end{eqnarray}
with 
\begin{eqnarray}
 c_{nn'} = \int c^2 (\omega) f_n (\omega) f_{n'}(\omega) d\omega .
\end{eqnarray}
Again, the $Q$-parameter is calculated as a matrix product of the variance $V$
multiplied from the left and from the right by a vector. This time, however, the
vector is not of the unit length, so that $Q^{f}$ is limited by
\begin{eqnarray}
 2 A^2 \left( V^{\text{(min)}}-\frac{1}{2} \right) 
 \leq Q^f \leq 2 A^2 \left( V^{\text{(max)}}-\frac{1}{2}\right) ,
\end{eqnarray} 
where $A^2$ is the square of the magnitude of the multiplying vector,
\begin{eqnarray}
 A^2 = \sum_{n\leq N} \left( \tilde X^{f2}_n + \tilde P^{f2}_n\right) .
\end{eqnarray}
To show that $Q^f$ can never be smaller than the minimum value achievable by
local oscillator modulation, it is enough to show that $A^2 \leq 1$, i.e., that
the magnitude of the multiplying vector is not bigger than one. Proof of this
inequality is shown in Appendix \ref{App-proof-vector}.

%%%%%%%%%%%%%%

\subsection{Conclusion}

We can see that no spectral filtering can improve the photon number squeezing
below the minimum eigenvalue of the quadrature variance matrix, which is 
available via the local oscillator modulation approach. Let us stress that our
method of finding the optimum local oscillator function is very simple and
straightforward since it is based on linear algebra. It was suggested recently
to use an adaptive algorithm to optimize the pulse shape for achieving optimum
photon number squeezing \cite{Takeoka}. In this method, the optimum was reached
after 20~000 iterations. In our case, provided that the pulse is sufficiently
well described by $N=5$ modes, $N(2N+1)=55$ iterations is enough.

%%%%%%%%%%%%%%%%%%%%%%%%%%%%%%%%%%%%%%%%%%%%%%%%%%%%%%%%%%%%%%%%%%%

\section{Selection of appropriate mode functions}
\label{S-modeselect}

So far it was assumed that the set of mode functions was given and the question
was about the statistics of the corresponding quadratures. But how should these
mode functions be selected?

In principle, any orthogonal set of mode functions can be used for description
of the pulse statistics. However, since the aim is to reduce the number of
quantum variables necessary to describe the pulse, the functions should be
carefully chosen. One possibility is to start from  theory and assume some
particular shape of the pulse - e.g., a hyperbolic secant soliton and to
construct the orthogonal set from the typical perturbations. An example  is
given in App. \ref{App-Haus}, Eqs. (\ref{nX1})--(\ref{pP24}) where relationship
to the soliton quantum fluctuations approach by Haus and Lai \cite{HausLai} is
discussed.

Another approach does not assume any particular mode function form and is
related directly to the experiment. It is useful to assume that only one mode
has a nonzero coherent amplitude and a very small number of modes have
quadrature fluctuations substantially different from the vacuum values. To find
the optimum set of mode functions, one can use the following procedure.

\begin{enumerate}

\item
Select $f_1(\omega)$ as the measured classical envelope of the pulse
(see App. \ref{S-Elimin}).
\label{item1}

\item
\label{item-minval}
Determine the minimum value of variance to be still considered as different from
vacuum.

\item
Construct a temporary set of orthogonal functions $f^{\text{(temp)}}_k$,
$k=1\dots N'$ 
with $f^{\text{(temp)}}_1= f_1$ chosen in \ref{item1}. 
The size $N'$ of the temporary
set should reasonably correspond to the experimental
conditions.

\item
Perform the measurement of the $2N'\times 2N'$
variance matrix $V^{(1)}$ with the temporary set of mode
functions.

\item
Construct the reduced matrix $V^{(2)}$ from $V^{(1)}$
by excluding rows and columns referring to quadratures of mode 1.
Let indexes $1\dots N'-1$ correspond to the $X$ quadratures, and indexes
$N'\dots N'-2$ correspond to the $P$ quadratures.

\item
\label{item-diag}
Diagonalize $V^{(2)}$ to get $\tilde V^{(2)} = WV^{(2)}W^{T}$ with
$W$ an orthogonal matrix. Let us choose $W$ such that $\tilde V^{
(2)}_{11}$ is the largest element of $\tilde V^{(2)}$; $\tilde V^{(2)}_{11} =
\sum_{k,l}W_{1k}W_{1l}V^{(2)}_{k,l}$.

\item
\label{item-f2}
The function $f_2$ is constructed as
\begin{eqnarray}
 f_{2} = \sum_{k=1}^{N'-1}\left( W_{1k} + i W_{1k+N'-1} \right) 
 f^{\text{(temp)}}_{(k+1)} .
\end{eqnarray}
It can be checked that the variance of the $X$ quadrature corresponding to this
mode function is $\langle \hat X_2^2 \rangle = \tilde V^{(2)}_{11}$.

\item
A new temporary set of $N'-2$ mode functions is constructed from the old one as
an orthogonal complement to $f_1$ and $f_2$. By means of the
transformation connecting the new temporary set of mode functions to the
initial one, calculate the corresponding variance matrix. Construct reduced
matrix $V^{(3)}$ by excluding rows and columns referring to quadratures  of
modes 1 and 2. Diagonalize $V^{(3)}$ and find mode function $f_{3}$ in the same
way as in \ref{item-diag} and \ref{item-f2}. Note that the variances of the
quadratures of mode 3 are smaller than the variance $\langle \hat X_2^2
\rangle$.

\item 
Repeat this procedure of redefining the temporary mode set, transforming and
diagonalizing the variance matrix, and defining a new mode function. After each
repetition, the maximum variance of the $(n+1)$st mode is smaller than
the maximum variance of the $n$th mode. If 
for some $n=N$ the variance 
of mode $n+1$ is sufficiently close
to $1/2$ (as defined in item \ref{item-minval}), 
the quantum state of the $(n+1)$st
mode is indistinguishable from vacuum. The pulse can be described 
with the given precision as an $N$-mode object.

\end{enumerate}

Let us note that the procedure of redefining the temporary mode set and
rediagonalizing the reduced variance matrix each time when a new mode function
is constructed is necessary. It is not possible, as one might be tempted, to
find a set of mode functions simply by diagonalizing the measured variance
matrix, since the SO(2$N$) transformation corresponding to diagonalization is
generally {\em not} a canonical transformation (see, e.g. \cite{SMD94} for more
details). On the other hand, in the special case of {\em pure\/} Gaussian states
which are specified by $N(N+1)$ real parameters one can find a set of
uncorrelated modes. In this case our procedure would be terminated after the
first diagonalization. The diagonalization of pure Gaussian multimode states
into uncorrelated modes has recently been studied in \cite{Bennink}.

The procedure of operational construction of mode functions as described above
is, of course, not the only possible. It is, however, very useful for finding
the minimum subspace of mode functions sufficient for the pulse description.
Other sets of mode functions can be selected such that the variance matrix
takes some special shape, e.g., some of the ``canonical'' forms studied in
\cite{SMD94}.

%%%%%%%%%%%%%%%%%%%%%%%%%%%%%%%%%%%%%%%%%%%%%%%%%%%%%%%%%%%%%%%%%%%

\section{Separation of modes}
\label{Sec-Separ}

It may be very useful to separate individual nonmonochromatic modes which form
the pulse. 
The possibility of obtaining nonclassical correlations of optical pulses
by partitioning the pulse in the spectral region was suggested in \cite{SKW00}.
Similarly as with spectral filtering, this approach is not necessarily the
optimum one for obtaining maximum entanglement from the source. 

\begin{figure}[htb]
\noindent
\centerline{\epsfig{figure=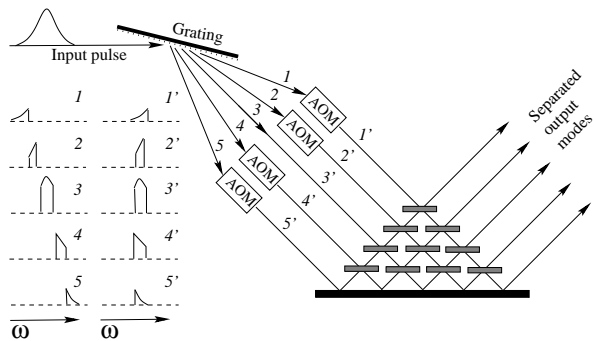,width=0.9\linewidth}}
\\[1ex]
\caption{ 
Scheme to separate individual nonmonochromatic modes. The pulse is
first decomposed on a grating into quasimonochromatic channels. 
The frequency of each channel is then shifted by means of 
an AOM to the central
frequency of the pulse. The resulting modes (with the same central frequency)
then interfere on a 2$N$-port.}
\label{f-aoms}
\end{figure}

The basic idea for obtaining the optimum separation
is to send different (nonmonochromatic) modes into
different channels by using a special unitary transformation among them. It was
shown in \cite{Reck} that any discrete unitary operator can be constructed
interferometrically. For this purpose we suggest to apply a scheme as in Fig.
\ref{f-aoms}. In the first step one decomposes the pulse into
quasimonochromatic components. The frequencies of these components are then
shifted by means of acusto-optical modulators (AOMs) so that each channel has
the same  central frequency. The channels then interfere on a 2$N$-port
consisting of  beam splitters and mirrors. By a proper choice of the 2$N$-port
parameters one can manage that most of a pulse in the $f_k(\omega)$ 
mode leaves the 2$N$-port   in the $k$th output channel.

%%%%%%%%%%%%%%%%%%  F I G U R E %%%%%%%%%%%%%%%%%%%%%%
%%%%%%%%%%%%%%%%%%%%%%%%%%%%%%%%%%%%%%%%%%%%%%%%%%%%%%%

One can also optimize the 2$N$-port parameters to prepare a set of
optimally entangled $K$ modes (multipartite entanglement), sent to different
channels. This would be a generalization of the proposal of \cite{SKW00} for
partitioning soliton pulses to generate entangled states.

Let us note that the spectral filtering of solitons used to produce photon
number squeezing \cite{Friberg} is a special kind of mode separation.  However,
as shown in Sec. \ref{S-comparison}, in this case the separation is not
optimized with respect to all relevant degrees of freedom. Therefore, better
results can be expected in our general approach.

%%%%%%%%%%%%%%%%%%%%%%%%%%%%%%%%%%%%%%%%%%%%%%%%%%%%%%%%%%%%%%%%%

\section{Discussion and conclusion}
\label{S-conclusion}

To summarize our results we can state that:

$(i)$ For a complete description of the experimental results measured on optical
solitons it appears that a very small number of nonmonochromatic modes is
enough. The results cannot be described by means of a single-mode field, but
already three nonmonochromatic modes were sufficient to reproduce the
experimental data of \cite{SKKSL98}. With our approach it is easy to interpret
the ``butterfly'' pattern of the measured photon covariances: three different
non-monochromatic modes overlap and contribute with their fluctuating
in-phase quadratures to the photon statistics (see Fig. \ref{f-modes}).

$(ii)$ Provided that the pulses are Gaussian (which seems to be a relevant
assumption for most of the experimental situations), a complete quantum
description of the pulse can be done by means of a multimode variance matrix.
The elements of this matrix can be determined by means of homodyne detection
with specially shaped local oscillator pulses.

$(iii)$ Concrete form of the mode functions is a matter of choice. If the aim
is to find the smallest number of modes, an operational method for constructing
the mode functions is provided. In this case the first mode contains the
coherent amplitude, whereas the rest of the modes have zero mean fields and
their field variances decrease with increasing mode index. 
By selecting the minimum set of modes, one can substantially reduce the number
of parameters necessary for a complete description of the physical situation.

$(iv)$ Knowledge of the mode structure of the pulse and of the corresponding
quantum state  will be very useful for quantum information purposes: one can
select the optimum shape of the local oscillator pulse to detect maximum
squeezing or entanglement.  This optimum finding is very straightforward and
much faster than  adaptive algorithms as in \cite{Takeoka}.
One can also relatively easily study the influence of the medium on the
propagated pulses and on the quantum information they carry. By measurement of
the multimode quantum state of the input and output pulses we can find the von
Neumann entropy of the states. This would enable us to tell whether the
observed pulse deformation corresponds most probably to a unitary evolution or
rather to decay and dephasing, possibly caused by an eavesdropper.

$(v)$ In principle, one can also separate individual modes
to match the requirements of the pulse user, e.g., to prepare a single mode
maximally squeezed field, or to extract a maximally entangled two-mode field,
etc. 

$(vi)$ The pulse separation is generalization of spectral filtering which has
also been used for observation of photon number squeezing \cite{Friberg}. We
have shown that spectral filtering can never beat the coherent amplitude
modulation approach in achieving better photon number squeezing. In our
approach, we can understand why spectral filtering can help with squeezing, but
we can also see its limitations.

$(vii)$ Our approach of multimode description of solitons is different from the
approach by Haus and Lai \cite{HausLai} (see App. \ref{App-Haus} for more
details). The formalism of \cite{HausLai} was developed to solve an idealized
quantum soliton equation and to study the soliton dynamics. Its essential
feature is the decomposition of the field into the ``soliton'' and
``continuum'' part, where the soliton field is completely described by four
operators. Our approach ignores the dynamics (which is rather complicated in
real life), focusing on the phenomenological description of the pulse. The
four soliton operators of \cite{HausLai} can be expressed by means of our
quadrature operators, but they alone appear to be insufficient for description
of the observed phenomena. It is necessary to work with the complete set
of ``soliton'' and ``continuum'' operators if one wishes to describe the
observed pulses using the formalism of \cite{HausLai}.

\acknowledgments

We are grateful to
R.S. Bennink,
R.W. Boyd,
J.~Fiur\'{a}\v{s}ek,
F.~K\"{o}nig, 
V.~Kozlov,
R.~Loudon,
A.~Matsko,
C.~Silberhorn, and
{D.--G.~Welsch}
for many stimulating discussions.
T.O. thanks to  Prof. G.~Leuchs for his 
kind hospitality at the Friedrich-Alexander
University in Erlangen.

%%%%%%%%%%%%%%%%%%%%%%%%%%%%%%%%%%%%%%%%%%%%%%%%%%%%%%%%%%%%%%%%%

\appendix

%%%%%%%%%%%%%%%%%%%%%%%%%%%%%%%%%%%%%%%%%%%%%%%%%%%%%%%%%%%%%%%%%%%%%%%%%%

\section{Simplified calculation of transformed mean quadratures and variances
for few excited modes} 
\label{App-simplify}

Let us assume that only the first $N$ modes of the $\hat b_k$ system are
occupied, while the rest is in vacuum, $|0\rangle _{k}$ for $k>N$. The
quadrature means and variances are thus
\begin{eqnarray}
 \overline{\mu}_k = 0 \quad {\rm for} \ 
 k>N ,
\end{eqnarray}
and 
\begin{eqnarray}
 V_{\mu k, \mu k} & = & \frac{1}{2} \quad \text{for} \ k > N , 
 \nonumber \\
 V_{\mu k, \mu' k'} &=& 0 \quad \text{for} \ \mu \neq \mu' 
 \nonumber \\
 & & \text{and} \
 (k>N \ \text{or} \ k' >N).
\end{eqnarray}
The transformations (\ref{trans2}) and (\ref{trans4}) can then be written as
\begin{eqnarray}
 \bar \xi _j =  \tilde{\sum_{\mu,k}} Z^{\mu \xi}_{k j}
 \bar{\mu}_k , 
 \label{simplxi} 
\end{eqnarray} 
and
\begin{eqnarray}
 V'_{\xi j,\xi' j'} =  
 \tilde{\sum_{\mu,k}} \tilde{\sum_{\mu,k'}}Z^{\mu\xi}_{k j}\ 
 V_{\mu k, \mu' k'}\ Z^{\mu'\xi'}_{k'j'} 
 \nonumber \\
  + \frac{1}{2} \sum_{\mu}\sum_{k>N}  Z^{\mu\xi}_{kj}\  
 Z^{\mu,\xi'}_{kj'} \
 \label{simplV0}
  \\
 \! = \!  \tilde{\sum_{\mu,\mu' k,k'}}  \!
  \left[ Z^{\mu \xi}_{kj} \left( 
  V_{\mu k,\mu'k'} \! - \! 
 \frac{\delta_{\mu \mu'} \delta_{kk'}}{2} \right) 
  Z^{\mu'\xi'}_{k'j'} \right] 
  %\nonumber \\
 \!  + \! \frac{\delta_{\xi \xi'}\delta_{jj'}}{2}
 \nonumber \\
  \label{simplV}
\end{eqnarray}
with
\begin{eqnarray}
 \tilde{\sum_{\mu,k}} \equiv \sum_{\mu=X,P}  \sum_{k=1}^N .
\end{eqnarray}
To get (\ref{simplV}) from (\ref{simplV0}), the orthogonality and completeness
relation (\ref{completZ}) was used.
As a special case, let us consider the variance matrix elements $V'_{xj,xj'}$
when the mode functions $f_{k}(\omega)$ are real,
as in Sec. \ref{Subsec-phot-stat}. Eq. (\ref{simplV}) then
becomes
\begin{eqnarray}
 V'_{xj,xj'} =  \frac{\delta_{jj'}}{2}
 \nonumber \\
 + \! \sum_{k,k' \leq N} f_{k}(\omega_{j}) f_{k'}(\omega_{j'})
 \Delta \omega \left( V_{Xk,Xk'} \! -\! \frac{\delta_{kk'}}{2} \right) .
 \label{simplVI}
\end{eqnarray}

%%%%%%%%%%%%%%%%%%%%%%%%%%%%%%%%%%%%%%%%%%%%%%%%%%%%%%%%%%%%%%%%%

\section{Quadrature moments of multimode Gaussian states}
\label{App-Gauss}

An $N$-mode Gaussian state
with mean quadratures  $\overline \mu$ and variance matrix $V$ has the Wigner
function $W(\mu)$,
\begin{eqnarray}
 W(\mu) \! =  \! \frac{1}{(2\pi)^N \! \sqrt{{\rm det}\ V}}
 \exp \left[ -\frac{  \left( \mu \!  -\!  \overline{\mu} \right) ^{T} 
 \!  V^{ -1} \! \left( \mu \!  - \! \overline{\mu} \right)  }
 {2}\right],
 \label{Wigner}
\end{eqnarray}
where $V^{-1}$ is the matrix inversion of $V$.
Thus, a Gaussian state is fully determined by 2$N^2$ $+$ 3$N$ real parameters
(2$N$ mean quadratures plus $N(2N+1)$ independent elements of the symmetric
matrix $V$). 

Generally, a quadrature moment $\langle \hat \mu_k^{n} \rangle$ 
can be calculated as
the integral of the Wigner function
\begin{eqnarray}
 \langle  \hat \mu_k^{n} \rangle = \int \dots \int  \mu_k^{n}   W(\mu)
 dX_1 \dots dP_{N} .
\end{eqnarray}
The symmetrical two-variable moments 
$\frac{1}{2}\left\langle \left\{ \hat \mu _k ,
 \hat \mu'_{k'} \right\}  \right\rangle$ 
 and $\frac{1}{2}\left\langle \left\{ \hat \mu_k^2 ,
 \hat \mu_{k'}^2 \right\}  \right\rangle$ can be calculated as
\begin{eqnarray}
 \frac{1}{2}\left\langle \left\{ \hat \mu_k ,
 \hat \mu'_{k'} \right\}  \right\rangle  = \int \dots \int  \mu_k 
   \mu'_{k'}  W(\mu)
 dX_1 \dots dP_N
\end{eqnarray}
and
\begin{eqnarray}
 \frac{1}{2} \! \left\langle \left\{ \hat \mu_k^2 ,
 \hat \mu_{k'}^{\prime 2} \right\}  \right\rangle    =   
 \int  \dots   \int  \mu_k^2   \mu_{k'}^{\prime 2}  W(\mu)
 dX_1 \dots dP_N   \nonumber \\
 -   \frac{1}{2}\tilde \delta_{\mu k,\mu' k'} , \! 
\end{eqnarray}
where $\tilde \delta_{\mu k,\mu' k'}$ $\equiv$ 0 if 
$\mu_k$ and $\mu'_{k'}$ commute with each other, and
$\tilde \delta_{\mu k,\mu' k'}$ $\equiv$ 1
if  $\mu_k$ and $\mu'_{k'}$ are conjugate variables.
Using Eq. (\ref{Wigner}) one can analytically evaluate the integrals and express
the quadrature
moments by means of the parameters  $\overline \mu_k$ and $V'_{\mu k, \mu' k'}$.
We obtain
\begin{eqnarray}
\langle  \hat \mu_k \rangle &=&  \overline{\mu}_k , 
\label{ximom1} \\
\langle  \hat \mu_k^2 \rangle &=&  \overline{\mu}_k^2 + V_{\mu k,\mu k} , 
\label{ximom2} \\
\langle  \hat \mu_k^3 \rangle &=&  \overline{\mu}_k^3 + 
  3\overline{\mu}_k V_{\mu k,\mu k} , 
\label{ximom3} \\
\langle  \hat \mu_k^4 \rangle &=&  \overline{\mu}_k^4 + 
  6\overline{\mu}_k^2 V_{\mu k,\mu k} + 3 V_{\mu k,\mu k}^2 , 
\label{ximom4} \\  
 \frac{1}{2}\left\langle \left\{ \hat \mu_k ,
 \hat \mu'_{k'} \right\}  \right\rangle  & = &
 \bar{\mu}_k \bar{\mu}_{k'} + V_{\mu k,\mu' k'} ,
\label{ximom5}  
\end{eqnarray}
and
\begin{eqnarray}
 \frac{1}{2}\left\langle \left\{ \hat \mu_k^2 ,
 \hat \mu_{k'}^{\prime 2} \right\}  \right\rangle  =   
 \bar{\mu_k}^2 \bar{\mu}_{k'}^{\prime 2} +  \bar{\mu_k}^2 V_{\mu' k',\mu' k'}
  +  \bar{\mu}_{k'}^{\prime 2} V_{\mu k,\mu k}   \nonumber \\
  + V_{\mu k, \mu k} V_{\mu' k', \mu' k'}
  + 4 \bar{\mu_k} \bar{\mu'}_{k'} V_{\mu k, \mu' k'} 
  \nonumber \\
  + 2 V_{\mu k, \mu' k'}^2 - \frac{1}{2}
  \tilde \delta_{\mu k, \mu' k'} .
\label{ximom6}  
\end{eqnarray}
Note that although (\ref{ximom1}),  (\ref{ximom2}), and 
(\ref{ximom5}) are generally valid for all states, the equalities 
(\ref{ximom3}), (\ref{ximom4}), and  (\ref{ximom6}) only hold for Gaussian
states. Corresponding expressions can be found also for the moments of the 
quadratures $\xi (\omega)$.

%%%%%%%%%%%%%%%%%%%%%%%%%%%%%%%%%%%%%%%%%%%%%%%%%%%%%%%%%%%%%%%%%%%%%%

\section{Photon statistics in different frequency channels of a single
nonmonochromatic mode}
\label{App-singlemode}

Let us assume a nonmonochromatic mode defined by the discrete function
$f(\omega_k)$, $k=1\dots N$,  $\sum_{k=1}^N|f(\omega_k)|^2$ $=$ 1.
Let the state of this mode be Gaussian. We are interested in the photon number
correlations between different frequency channels.

{\bf Theorem:} All non-zero elements of a multi-mode  
photon correlation matrix of an
effectively single-mode  Gaussian state have the same sign. They are 
negative (positive) iff the single-mode state
is sub(super)-Poissonian.

{\bf Proof:} The mode function $f(\omega_k)$ defines the first row of the unitary
transformation matrix $U$. Let the parameters of the single-mode
Gaussian state be $\bar X$,  $\bar P$, $V_{XX}$,  $V_{PP}$, and $V_{XP}$.
The multi-mode parameters $\bar x_{k}$,  $\bar p_{k}$,
$V_{xk,xl}$,  $V_{xk,pl}$, etc., can be calculated by means of the simplified
summation as in (\ref{simplxi}) and (\ref{simplV}).
The sign of the off diagonal element of the correlation matrix
(\ref{correl-off}) is determined by the sign of the covariance (\ref{covar}).
After some algebra with applying Eqs. 
(\ref{simplxi}), (\ref{simplV}), and  (\ref{covarnxp}) we arrive at
\begin{eqnarray}
 {\rm cov}\ (n_k,n_l) = |f(\omega _k)|^2|f(\omega _l)|^2 \nonumber \\
 \times \left[ \bar X^2 
 \left( V_{XX} \! -\! \frac{1}{2} \right)
 +\ 2 \bar X \bar P V_{XP} +
 \bar P^2  \left( V_{PP} \! -\! \frac{1}{2} \right) \right.  \nonumber \\
 \left.
 + \frac{1}{2} \left( V_{XX} \! -\! \frac{1}{2} \right) ^2
 + V^{\prime 2}_{XP} +
  \frac{1}{2} \left( V_{PP} \! -\! \frac{1}{2} \right)^2 \right] .
\end{eqnarray}
Thus, we can see that the sign of the non-zero elements does not depend on the
arguments $\omega_k$ and $\omega_l$, but is fully determined by the expression
in the square brackets. If we denote the photon number in the single mode by
$\hat n$, we find that the quantity $\Delta n^2 - \langle n \rangle$ is equal
to the expression in the square brackets. This quantity is negative for
sub-Poissonian states and positive for super-Poissonian states by definition, 
QED.

%%%%%%%%%%%%%%%%%%%%%%%%%%%%%%%%%%%%%%%%%%%%%%%%%%%%%%%%%%%%%%%%%

\section{Proof that the magnitude of the 
spectral-filtering quadrature vector is less than 1}

\label{App-proof-vector}

Showing that 
\begin{eqnarray}
 A^2 = \sum_{n\leq N} \left( \tilde X^{f2}_n + \tilde P^{f2}_n\right) 
 \leq 1
\end{eqnarray}
is equivalent to showing that
\begin{eqnarray}
 \sum_{k,l,n\leq N}  c_{kn} c_{ln} 
 \left( \bar X_k \bar X_l + \bar P_k  \bar P_l \right)
 \nonumber \\
 \leq \sum_{k,l \leq N} c_{kl} (\bar X_k \bar X_l + \bar P_k \bar P_l),
 \label{E-vecproof1}
\end{eqnarray}
which follows directly from the definitions of $\tilde X^f$ and $\tilde P^f$,
Eqs. (\ref{tildeXf}), (\ref{tildePf}).
To violate Eq. (\ref{E-vecproof1}), it would be necessary to have some vector $Y$
such that 
\begin{eqnarray}
 Y^T \underline{c}^2  Y > Y^T \underline{c} Y ,
 \label{E-vecproof2}
\end{eqnarray}
where $\underline{c}$ is a matrix with the elements $c_{kl}$.
Eq. (\ref{E-vecproof2}) could only be valid if there exists some eigenvalue 
$c_{\lambda}$ of $\underline{c}$ such that $c_{\lambda} > 1$. Let us assume that
such an eigenvalue does exist. Let $u_k$ be the elements of the corresponding
eigenvector, i.e.,
\begin{eqnarray}
 \sum_{l\leq N} c_{kl} u_{l} = c_{\lambda} u_{k},
\end{eqnarray}
i.e., 
\begin{eqnarray}
 \sum_{l \leq N} u_l \int c^2(\omega ) f_{k}(\omega) f_{l}(\omega) d\omega
 = c_{\lambda} u_k.
\end{eqnarray}
Let us define a function $q(\omega)$ as
\begin{eqnarray}
 q(\omega) = \sum_{l\leq N}u_l f_l(\omega) .
\end{eqnarray}
Since $u_k = \int q(\omega) f_k (\omega) d\omega$, one has
\begin{eqnarray}
 \int c^2(\omega) q(\omega) f_k(\omega)d\omega = c_{\lambda}
 \int q(\omega) f_k (\omega) d\omega ,
\end{eqnarray}
i.e.,
\begin{eqnarray}
 \left| \int c^2(\omega) q(\omega) f_k(\omega) d\omega \right|
 > \left| \int q(\omega) f_k(\omega) d\omega \right| ,
\end{eqnarray}
which cannot happen for any function $q(\omega)$ since $|c^2(\omega)| \leq 1$.
Thus, $A^2 \leq 1$, QED.

%%%%%%%%%%%%%%%%%%%%%%%%%%%%%%%%%%%%%%%%%%%%%%%%%%%%%%%%%%%%%%%%

\section{Elimination of coherent amplitudes of a multi-mode Gaussian state}
\label{S-Elimin}

Let us assume that $N$ modes of the $b$-system are excited and the rest is in
vacuum. Then, one can redefine the $N$ modes such that only one of them  (say
mode $\hat b_1$) has nonzero values of $\bar X_1$, $\bar P_1$, whereas  $\bar
X_m$ $=$ 0,  $\bar P_m$ $=$ 0 for $m$ $>$ 1 (the variance matrix elements
corresponding to these modes are, however, generally non-zero).

{\bf Proof:} Define $\beta_m$ $\equiv$ $2^{-1/2}(\bar X_m + i \bar P_m)$ $=$
$\langle \hat b_m \rangle$, $m$ $=$ $1\dots N$. Let $f_m$ be the mode functions
normalized such that $(f_m,f_{m'})$ 
$=$ $\delta_{mm'}$, where the bracket denotes
the scalar product. 
Let us define a new
mode function $g_1$ as $g_1$ $\equiv$ $A \sum_{m=1}^{N}\beta^{*}_m f_m$ and a
corresponding annihilation operator $\hat h_1$ $\equiv$    
$A \sum_{m=1}^{N}\beta^{*}_m \hat b_m$, where $A$ $=$ $\left( \sum_{m=1}^{N}
|\beta_{m}|^{2}\right)^{-1/2}$. Then $\langle \hat h_1 \rangle$ $=$
$\sqrt{\sum_{m=1}^{N}|\beta_{m}|^{2}}$. Let us define mode functions 
$g_2 \dots g_N$ as linear combinations of $f_1 \dots f_N$ by some
orthogonalization procedure, i.e., $g_m$ $=$ $\sum_{m'=1}^N G_{mm'} f_{m'}$,
$m=2\dots N$ such that $(g_m,g_{m'})$ $=$ $\delta_{mm'}$, $m=1\dots N$. In
particular, 
\begin{eqnarray}
 (g_1,g_m) = A \sum_{l=1}^{N}\beta_l G_{ml} = 0
 \label{orthog}
\end{eqnarray}
for $m>1$. The annihilation operators corresponding to the $g$-modes are
$\hat h_m$ $\equiv$ $\sum_{m'=1}^N G_{mm'} \hat b_{m'}$, and their mean values
are $\langle \hat h_m \rangle$ $=$ $\sum_{m'=1}^N G_{mm'} \langle \hat
b_{m'}\rangle$  $=$ $\sum_{m'=1}^N G_{mm'} \beta_{m'}$ $=$ 0 according to
(\ref{orthog}). Thus, in the new system of modes only the first one has a
non-zero coherent amplitude, QED.

%%%%%%%%%%%%%%%%%%%%%%%%%%%%%%%%%%%%%%%%%%%%%%%%%%%%%%%%%%%%%%%%%

\section{Relationship to the soliton perturbation approach
by Haus and Lai}
\label{App-Haus} 

In \cite{HausLai} (see also \cite{Haus00,MK00}) it is suggested to describe the
quantum fluctuations of the Nonlinear Schr\"{o}dinger equation 
(NSE) soliton by
writing
\begin{eqnarray}
 \hat a(x,t)=a_o(x,t) + \Delta \hat a (x,t),
\end{eqnarray}
where $a_o (x)$ is the unperturbed solution of the NSE of the hyperbolic secant
form, and  
\begin{eqnarray}
 \Delta \hat a = \Delta \hat a_{\text{sol}}
 + \Delta \hat a_{\text{cont}}
\end{eqnarray}
is the quantum perturbation part with
$\Delta \hat a_{\text{sol}}$ describing fluctuations of the soliton degrees of
freedom, while $\Delta \hat a_{\text{cont}}$ corresponds to field fluctuations
not contained in the soliton solution and thus belonging to the continuum.
The soliton fluctuations are expressed 
by means of four operators $\Delta \hat n$, $\Delta \hat \theta$,
$\Delta \hat x$, and $\Delta \hat p$ as
\begin{eqnarray}
 \Delta \hat a_{\text{sol}} = 
 \left[ \Delta \hat n(t) f_n(x) + \Delta \hat \theta (t) f_{\theta} (x)
 \right. \nonumber \\
 \left.
 + \Delta \hat x (t) f_{x}(x) + n_o \Delta \hat p (t) f_p (x) \right]
 \times \exp \left( i\frac{KA_o^2}{2} t \right),
\end{eqnarray}
where the functions $f_n$, $f_{\theta}$, $f_x$, and $f_p$ are derivatives 
of the soliton function
\begin{eqnarray}
 a_o(x,t) = A_o \exp \left[ i \left( \frac{KA_o^2}{2}t - \frac{C}{2}p_o^2t
 + p_o x + \theta_o \right) \right] \nonumber \\
 \times \text{sech} \left( \frac{x - x_o - Cp_o t}{\xi} \right)
\end{eqnarray}
with respect to the parameters $n_o (= 2\xi |A_o|^2)$, $p_o$, $\theta_o$, 
and $x_o$. In these equations
the value $n_o$ is the mean photon number of the pulse, $\xi$ is the
soliton width, $p_o$ and $x_o$ refer to the carrier frequency and soliton center
position, respectively. The value $K$ is related to the Kerr nonlinearity and
$C$ to the medium dispersion.
To find the values of the operators from measured data, one 
introduces adjoint functions $\underline{f}_n$, $\underline{f}_{\theta}$, 
$\underline{f}_x$, and $\underline{f}_p$ with the property 
\begin{eqnarray}
 \text{Re}\ \left[ \int \underline{f}^*_k(x) f_l(x) dx \right]
= \delta_{kl}
\end{eqnarray}
with $k,l = n,\theta, x, p$ (note that $f_k$ themselves do not form an 
orthogonal set; 
explicit form of functions $f_k$ and $\underline{f}_k$ 
can be found in \cite{Haus00}). Writing $\Delta \hat a$ as a sum of hermitian
operators $\Delta \hat a = \Delta \hat a ^{(1)} + i \Delta \hat a^{(2)}$ one
finds that
\begin{eqnarray}
 \Delta \hat n &=& \int \underline{f}_n^*(x) \Delta \hat a ^{(1)} (x) dx, \\
 \Delta \hat \theta &=& i \int 
 \underline{f}_{\theta}^*(x) \Delta \hat a ^{(2)} (x) dx, \\
 \Delta \hat x &=& \int \underline{f}_x^*(x) \Delta \hat a ^{(1)} (x) dx, \\
 \Delta \hat p &=& i \frac{1}{n_o}\int 
 \underline{f}_{p}^*(x) \Delta \hat a ^{(2)} (x) dx. \\ 
\end{eqnarray}

The relationship of this approach to our scheme can be easily examined if one
assumes the first four mode functions $f_k(\omega)$ in the form
\begin{eqnarray}
 \label{f1}
 f_1(\omega) &=& \frac{1}{\sqrt{2 \omega_o}} \text{sech} 
 \frac{\omega}{\omega_o} , \\
 f_2(\omega) &=& \sqrt{\frac{3}{2\omega_o}} 
 \text{tanh} \frac{\omega}{\omega_o} \text{sech} 
 \frac{\omega}{\omega_o} , \\ 
 \label{f3}
 f_3(\omega) &=& \frac{1}{\sqrt{\frac{1}{3}+\frac{\pi^2}{9}}}
 \frac{1}{\sqrt{2 \omega_o}} \nonumber \\
 &\times& \left( 2 \frac{\omega}{\omega_o}
 \text{tanh} \frac{\omega}{\omega_o} \text{sech} 
 \frac{\omega}{\omega_o} - \text{sech} \frac{\omega}{\omega_o} \right), \\
 f_4(\omega) &=& i \frac{\sqrt{3}}{\sqrt{\frac{\pi^2}{9}-1}}
 \frac{1}{\sqrt{2 \omega_o}} \nonumber \\
 &\times& \left(
 \text{tanh} \frac{\omega}{\omega_o} \text{sech} 
 \frac{\omega}{\omega_o}
 - \frac{2}{3} \frac{\omega}{\omega_o} \text{sech} \frac{\omega}{\omega_o}
 \right). 
\end{eqnarray}
These functions were obtained by an orthogonalization procedure from the Fourier
transformed functions $f_{n,\theta,x,p}$ with $\omega_{o}= 2c/(\pi \xi)$. With
this choice of the mode functions one finds that the soliton perturbation
operators can be expressed by means of the quadratures $\hat X_1$, $\hat X_2$,
$\hat P_1$, $\hat P_2$, $\hat P_3$, and $\hat P_4$ as
\begin{eqnarray}
 \label{nX1}
 \Delta \hat n &=& \sqrt{2 n_o} \hat X_1 , \\
 \label{thP13}
 \Delta \hat \theta &=& \frac{1}{\sqrt{2n_o}} \hat P_1 + 
 \frac{\sqrt{\frac{1}{3}+\frac{\pi^2}{9}}}{\sqrt{2n_o}} \hat P_3 , \\
 \label{xX2}
 \Delta \hat x &=& \frac{2c}{\sqrt{6 n_o}\omega_o} \hat X_2 , \\
 \label{pP24}
 \Delta \hat p &=& \frac{\sqrt{6n_o}\omega_o}{2c} \hat P_2 
 - \sqrt{\frac{\pi^2}{9}-1} \frac{\sqrt{6n_o}\omega_o}{2c} \hat P_4.
\end{eqnarray}
Assuming vacuum fluctuations of $\hat X_k$
and $\hat P_k$, $\langle X_k^2 \rangle = 1/2$, $\langle P_k^2 \rangle = 1/2$, one
finds that the uncertainty products of the soliton perturbation operators are
\begin{eqnarray}
 \langle \Delta \hat n^2 \rangle \langle \hat \Delta \theta ^2 \rangle &=&
 \frac{1}{3} + \frac{\pi^2}{36} \approx 0.675, \\
 \langle \Delta \hat x^2 \rangle \langle n_o^2 \Delta \hat  p^2 \rangle &=&
 \frac{\pi^2}{36} \approx 0.274,
\end{eqnarray}
which are larger than the minimum uncertainty value $1/4$ following from the
commutation relations $[\Delta \hat n,\Delta \hat \theta] = i$ and
$[\Delta \hat x, n_o \Delta \hat p] = i$. This result corresponds exactly to
that of \cite{HausLai,Haus00}. Our interpretation of the result is that the
operator $\Delta \hat \theta$ is not purely
a conjugate of $\Delta \hat n$, but it
contains an admixture of an operator commuting with $\Delta \hat n$.
This admixture [in (\ref{thP13}) proportional to $\hat P_3$] increases the noise
above the minimum uncertainty limit. Similar interpretation holds 
for the pair $\Delta \hat x$, $n_o \Delta \hat p$.

{F}rom Eqs. (\ref{nX1})--(\ref{pP24}) one sees that from the statistics of $\hat
X_{1,2}$ and $\hat P_{1,2,3,4}$ one can determine the statistics of the soliton
perturbation operators
$\Delta \hat n$, $\Delta \hat \theta$, $\Delta \hat x$ and $\Delta \hat p$. This
does not hold vice versa: knowledge of the four soliton operators is not enough
to determine the six quadratures.

%%%%%%%%%%%%%%%%%%%%%%%%%%%%%%%%%%%%%%%%%%%%%%%%%%%%%%%%%%%%%%%%%

\end{document}